\documentclass[9pt,twocolumn,twoside]{article}

\usepackage[utf8]{inputenc}
\usepackage[T1]{fontenc}
\usepackage{graphicx}
\usepackage{units}
\usepackage{dsfont}
\usepackage{upgreek}
\usepackage{amsfonts}
\usepackage{amsmath}
\usepackage{authblk}
\allowdisplaybreaks

\begin{document}

\title{Dynamics of the optical swallowtail catastrophe}

\author[]{A. Zannotti}
\author[]{F. Diebel}
\author[]{C. Denz}
\affil[]{Institute of Applied Physics and Center for Nonlinear Science (CeNoS), University of Münster, 48149 Münster, Germany}
\affil[]{\textit {a.zannotti@uni-muenster.de}}
\date{April 2017}
\maketitle

\begin{abstract}
Perturbing the external control parameters of nonlinear systems leads to dramatic changes of its bifurcations. A branch of singular theory, the catastrophe theory, analyses the generating function that depends on state and control parameters. It predicts the formation of bifurcations as geometrically stable structures and categorizes them hierarchically. We evaluate the catastrophe diffraction integral with respect to two-dimensional cross-sections through the control parameter space and thus transfer these bifurcations to optics, where they manifest as caustics in transverse light fields. 
For all optical catastrophes that depend on a single state parameter, we analytically derive a universal expression for the propagation of all corresponding caustic beams. We reveal that the dynamics of the resulting caustics can be expressed by higher-order optical catastrophes. We show analytically and experimentally that particular swallowtail beams dynamically transform to higher-order butterfly caustics, whereas other swallowtail beams decay to lower-order cusp catastrophes. 
\end{abstract}

\section{Introduction}

Singularities of the generating function of a nonlinear system describe how the structures of its bifurcations are related to external control parameters. Small perturbations of the control parameters may result in dramatic changes. This is what is called a catastrophe in the context of singularity and catastrophe theory~\cite{Thom1975, Arnold2003}. A classification of the most fundamental potentials distinguishes different local effects near catastrophes and defines the system's qualitative behaviour~\cite{Thom1975, Arnold2003}. Each order of a catastrophe forms a specific geometrically stable structure in control parameter space. According to the unique geometric structure, a hierarchical categorization with increasing dimensionality of the control parameter space is given by the fold, cusp, swallowtail, hyperbolic umbilic, elliptic umbilic, butterfly, and parabolic umbilic catastrophe.

In general, any catastrophe has a wide range of different areas of physics that it explains~\cite{Berry1976, Gifford2013, Hoehmann2010, Mathis2015}, where individual catastrophes stand for miscellaneous systems. Exemplary, in meteorology the fold catastrophe manifests in rainbows~\cite{Trinkhaus1977}, the cusp catastrophe emerges in models of social sciences describing an employee turnover~\cite{Sheridan1983}, and the swallowtail catastrophe occurs in orbits of hydrogen in atom physics~\cite{Mendez2013}, to name only some exemplary orders of catastrophes. 

The manifestation of catastrophes in natural light phenomena is given by high-intensity caustics that appear in defined geometries and different orders~\cite{Berry1979, Berry1980}. Caustics form due to reflections on smoothly curved surfaces and emerge e.g. as a cusp in a cup, or arise due to refraction at spatially modulated boundaries e.g. when forming ramified networks at the ground of shallow waters. Beneath their potential to optically visualize the complex dynamics of nonlinear systems, it is their sharp high-intensity boundary with unique propagation paths on curved trajectories~\cite{Siviloglou2007a, Ring2012, Zannotti2017a} that makes caustics in light highly attractive for a broad range of applications~\cite{Diebel2015, Baumgartl2008, Manousidaki2016, Zannotti2017a}. However, the tailored creation of caustics in light beyond their natural occurrence is necessary to control their properties and dynamics in all facets. 

Only recently, the application of spatial light modulators has facilitated the artificial creation of catastrophes in light fields, but until now only the most fundamental orders were realized as paraxial Airy (fold) and Pearcey (cusp) beams~\cite{Siviloglou2007a, Ring2012}. Artificial higher-order optical swallowtail and butterfly catastrophes were transferred to caustics in transverse light fields by mapping cross-sections of the higher-dimensional control parameter space to the two-dimensional (2D) transverse plane~\cite{Zannotti2017}. The presented approach allows realizing fundamentally different geometrical caustic structures in the initial transverse plane without propagating the light fields by choosing corresponding cross-sections in the control parameter space. Their dynamics, however, are unknown. Of particular interest is the stability of higher-order caustic structures during propagation. They are expected to decay to lower-order caustics, as it is typical for high-dimensional singularities, since the initial caustic structures represent mappings of the related higher-order catastrophes to the transverse plane. Furthermore, the propagation of higher-order caustic beams is expected to provide similar breakthrough potential as their already well-known mates of lower order. 

Thus, with this work we provide for the first time to our knowledge the general analysis of the propagation of optical catastrophes mapped to transverse light fields by evaluating the catastrophe diffraction integral for a potential that depends on a single state parameter. We prove our universal findings by demonstrating optical swallowtail beams, a particular solution of the paraxial Helmholtz equation. We analytically calculate and prove experimentally how the higher-order swallowtail beams evolve in space. We show that the dynamics of the swallowtail beams can again be described in terms of swallowtail beams or contain fingerprints of the optical butterfly catastrophes, where the control parameters depend on $z$. For a different set of parameters, we demonstrate how the corresponding initial swallowtail caustic decays into a lower-order cusp during propagation.

\section{The propagation of optical catastrophes}

We consider the complete class of caustic beams emerging as optical catastrophes that depend on a single state parameter $s$. It represents a particular solution of the paraxial Helmholtz equation and emerges as solution of the diffraction catastrophe integral $C_n(\mathbf{a})$~\cite{Berry1980, Nye1999}
\begin{equation}
C_n(\mathbf{a}) = \int\limits_{\mathds{R}}{e^{\text{i}P_n\left(\mathbf{a},s\right)}}\mathrm{d}s, \qquad P_n\left(\mathbf{a},s\right) = s^n + \sum\limits_{j=1}^{n-2} \frac{a_j}{a_{0j}}s^j.
\label{eq:CatastropheIntegral}
\end{equation}
All resulting light structures $C_n(\mathbf{a})$ exhibit individual caustic profiles that are characterized by degenerate critical points of the potential function $P_n\left(\mathbf{a},s\right)$. Here, 
the vector $\mathbf{a}$ consists of all control parameters $a_j$ with $j = 1,...,n-2$ and spans the \textit{control parameter space} with co-dimension $n-2$. We identify two of the $n-2$ control parameters with transverse spatial coordinates~\cite{Berry1980} and keep the remaining $n-4$ control parameters constant. Dimensionless parameters are ensured by introducing characteristic structure sizes $a_{0j}$. We restrict the following discussion to optical catastrophes with $n \geq 4$.

We emphasize that the Airy beam $\text{Ai}(x) = C_3(\mathbf{a})$ results as 1D structure identified with the spatial coordinate $\mathbf{a} = a_1 = x$. Thus, our approach is also valid for $n = 3$ by reducing all transverse 2D considerations to 1D. However, the Pearcey beam $\text{Pe}(x, y) = C_4(\mathbf{a})$ leads per se to a 2D distribution, by identifying $\mathbf{a} = (a_1, a_2)^T = (x, y)^T$. Consequently, for higher-order optical catastrophes as e.g.~the swallowtail beams $\text{Sw}(\mathbf{a}) = C_5(\mathbf{a})$, we have to chose a single control parameter out of three existing ones $\mathbf{a} = \left(a_1, a_2, a_3\right)^T$ to be constant in order to map characteristics of the corresponding catastrophe to the 2D transverse light field by identifying the two remaining control parameters with spatial coordinates $(x, y)^T$. Thus, for the optical swallowtail catastrophe three generic swallowtail beams arise as orthogonal cross-sections through the control parameter space, and correspondingly more for the butterfly beam $\text{Bu}(\mathbf{a}) = C_6(\mathbf{a})$ that maps characteristics of the butterfly catastrophe. This approach is thoroughly presented in~\cite{Zannotti2017}.

Since caustics represent geometrically stable structures determined by the singular mapping of the potential function $P_n$ on the $n-2$ dimensional plane $\mathbf{a}$, we calculate degenerate fixed points $s_i$ as where the first and second derivatives vanish. The submanifold $s_i$ then subsequently determines the geometric structure of the catastrophe.
\begin{equation}
\left. \frac{\partial P_n}{\partial s} \right|_{s = s_i} = 0, \quad \text{and} \quad \left. \frac{\partial^2 P_n}{\partial s^2} \right|_{s = s_i} = 0.
\label{eq:DegeneratedCriticalPoints}
\end{equation}
By transferring these potentials to optics via \eqref{eq:CatastropheIntegral}, their singularities correspond to geometrically stable catastrophes and manifest as caustics in paraxial wave structures.

Starting with this formalism to create the class of paraxial caustic beams including the well known Airy and Pearcey beams as well as the less-known swallowtail and butterfly beams, we derive analytical expressions for the propagation of these intriguing light structures and furthermore analyze the $z$-dependent evolution of their caustics in the regime of paraxial light by referring to \eqref{eq:DegeneratedCriticalPoints}.

In order to analytically calculate the propagation of a scalar transverse light field, we apply the angular spectrum method~\cite{Goodman2005} and derive $z$-dependent expressions for caustic beams \eqref{eq:CatastropheIntegral} in the paraxial regime. That is, the propagation of a known transverse electric field distribution $C_n(\mathbf{a})$ can be determined by first calculating the 2D Fourier transform $\tilde{C}\left(k_\alpha,k_\beta,\mathbf{a}'\right)$ with respect to two control parameters $a_\alpha, a_\beta$ identified with $x, y$, where $\alpha, \beta \in j = 1,...,n-2$ and $k_\alpha, k_\beta$ are corresponding Fourier frequencies. We denote the remaining (constant) control parameters with $\mathbf{a}'$.  The general Fourier transform of \eqref{eq:CatastropheIntegral} can be found in the supplementary material (SM). 

Subsequently, we apply the Fresnel propagator $G\left(a_\alpha,a_\beta,k_\alpha,k_\beta,z\right)=\exp\left[\text{i}\left(k_\alpha a_\alpha + k_\beta a_\beta - \frac{z k_\alpha^2}{2 k} - \frac{z k_\beta^2}{2 k}\right)\right]$ in an integral expression. Here, $k = 2\pi/\lambda = \sqrt{k_\alpha^2 + k_\beta^2 + k_z^2}$ is the wave number and $\lambda$ the wavelength. Consequently, the $z$-dependence of the caustic beams is derived to be
\begin{eqnarray}
C_n(\mathbf{a},z) & = &\int\limits_{\mathds{R}^2} \tilde{C}_n\left(k_\alpha,k_\beta,\mathbf{a}'\right) G\left(a_\alpha,a_\beta,k_\alpha,k_\beta,z\right) \mathrm{d}k_\alpha \mathrm{d}k_\beta \nonumber \\
& = &\int\limits_{\mathds{R}} e^{\text{i}P_n\left(\mathbf{a},s\right)} e^{-\text{i}\frac{z}{z_{e\alpha}}s^{2\alpha}} e^{-\text{i}\frac{z}{z_{e\beta}}s^{2\beta}} \mathrm{d}s\:.
\label{eq:CausticPropagation}
\end{eqnarray}
Similar to the Rayleigh length of Gaussian beams, we can define characteristic Rayleigh lengths $z_{ex} = 2kx_0^2$ and $z_{ey} = 2ky_0^2$ that play an important role for the dynamics of the beams and depend on the wavelength $\lambda = 2\pi/k$ and the corresponding the structure sizes $x_0, y_0$. In general, $\mathbf{z_e} = (z_{e\alpha},z_{e\beta})^T = 2k(a_{0\alpha}^2,a_{0\beta}^2)^T$.

Using Eq.~$\left(\ref{eq:CausticPropagation}\right)$ with the proper choice of parameters, one can easily derive the well known propagation expressions for the Airy~\cite{Berry1979a} and Pearcey beam~\cite{Ring2012}, respectively. 

Eq.~$\left(\ref{eq:CausticPropagation}\right)$ reveals fundamentally new insights in the dynamics of caustics in light: In the case of the Airy and the Pearcey beams, their propagation is again expressed by Airy and Pearcey functions (displacement or form-invariant scaling, respectively)~\cite{Siviloglou2007a, Ring2012}. Starting with the order $n\geq 5$ with co-dimension $n-2$, the propagation of the corresponding caustic beam can be described in terms of caustic beams up to order $2(n-2)$, if identifying the appropriate control parameters with transverse spatial coordinates. This relation is manifested in the propagator $T_{\alpha, \beta}(s,z) = \exp\left[-\text{i}\frac{z}{z_{e\alpha}}s^{2\alpha}\right] \exp\left[-\text{i}\frac{z}{z_{e\beta}}s^{2\beta}\right]$ of \eqref{eq:CausticPropagation}, where one of the exponents $2\alpha$ or $2\beta$ can always be chosen to be larger then the leading exponent $n$ of the potential function $P_n$, and contributes only if $z \neq 0$. Thus, by choosing $\alpha$ or $\beta$ to be the coefficient that corresponds to the state variable of highest degree (i.e. $n-2$), the propagation of this $n^\text{th}$-order caustic beam is given by a static caustic beam with $2(n-2)$-dimensional control parameter space and $z$-depending control parameters.

In the following, we exemplary demonstrate our concept by applying \eqref{eq:CausticPropagation} for two initial transverse swallowtail beams to show their fundamentally different dynamics. First, we calculate and experimentally obtain the dynamics of a $\text{Sw}(x,y,a_3)$ beam. We show that its propagation can be described by an optical catastrophe of the same order with varying control parameters. Second, the analytical and experimental investigation of a $\text{Sw}(x,a_2,y)$ beam reveals that its dynamics are represented by a higher-order optical butterfly catastrophe whose control parameters are functions of $x,y,z,a_2$.

\begin{figure}[h]
\centering
\includegraphics[width=\columnwidth]{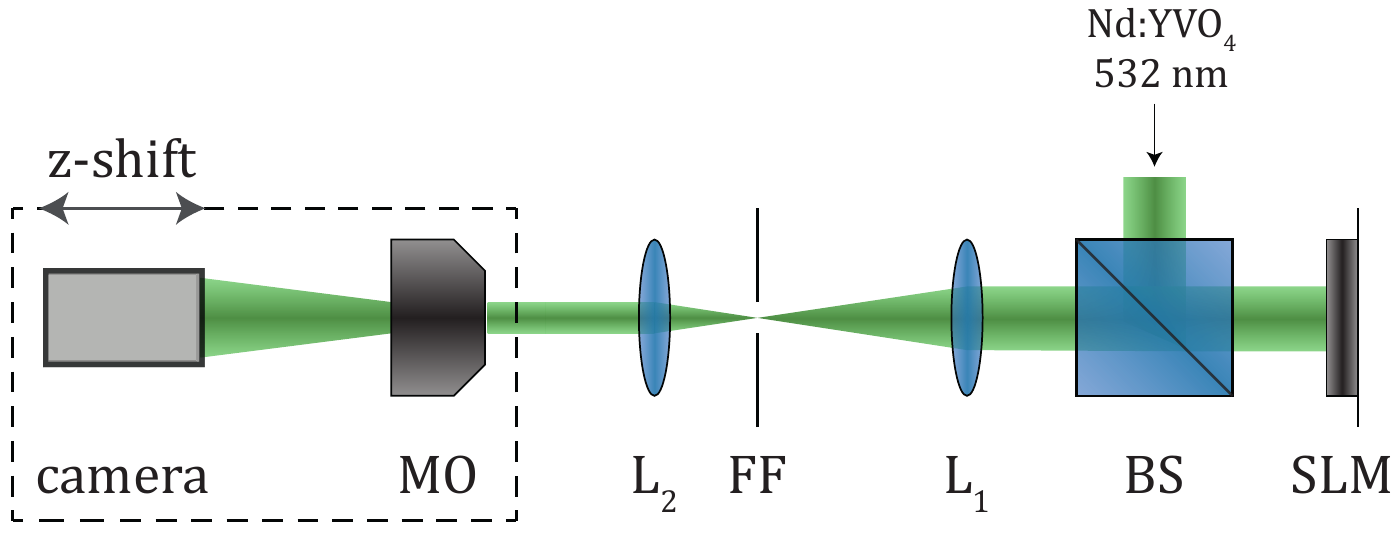}%
\caption{Scheme of experimental setup. BS: beam splitter, FF: Fourier filter, L: lens, MO: microscope objective, SLM: phase-only spatial light modulator.}%
\label{fig:Setup}%
\end{figure}

In order to obtain the spatial intensity distribution experimentally we have used the setup shown in Fig.~\ref{fig:Setup}. As light source serves a frequency-doubled Nd:YVO$_4$ laser with a wavelength of $\lambda = \unit[532]{nm}$. The linearly polarized beam is expanded and collimated. As plane wave, it illuminates a HOLOEYE HEO 1080P reflective LCOS phase-only spatial light modulator (SLM) with full HD resolution. We modulate both, amplitude and phase of the beam by encoding both informations in a phase-only pattern and applying an appropriate Fourier filter~\cite{Davis1999}. The beam is then imaged by a camera after passing a microscope objective. Both, camera and microscope objective are mounted on a in $z$-direction movable stage capable to scan the propagation of the light fields by obtaining transverse intensity patterns.

\section{Dynamics of a swallowtail beam}
\label{sc:SwPropagation}

As pointed out in our analytic description, particular emphasis will be on the difference between the dynamics of a $\text{Sw}\left(x,y,a_3\right)$ beam in comparison to these of the $\text{Sw}\left(x,a_2,y\right)$ or $\text{Sw}\left(a_1,x,y\right)$ beams, since the propagation of the $\text{Sw}\left(x,y,a_3\right)$ beam is a mapping from the three-dimensional (3D) control parameter space onto itself, whereas the evolution of the two other beams is described in the higher four-dimensional (4D) control parameter space. In order to investigate these dynamics, this section substantiates our analytical investigations by experimental realizations of the $\text{Sw}\left(x,y,a_3\right)$ beam and, arbitrarily chosen, the $\text{Sw}\left(x,a_2,y\right)$ beam. The dynamics of the $\text{Sw}\left(a_1,x,y\right)$ beam are similar to those of the $\text{Sw}\left(x,a_2,y\right)$ beam, as stated in \eqref{eq:CausticPropagation}, and are not shown here explicitly. The analytically calculated propagation is given in the SM.

For the $\text{Sw}\left(x,y,a_3\right)$ beam with $n = 5$ we chose $\alpha = 1$ and $\beta = 2$. \eqref{eq:CausticPropagation} then leads to a non-canonical $5^\text{th}$-order potential function in the exponential which can be brought in canonical form by applying the Tschirnhaus transform~\cite{Weisstein2008}. The dynamics of the $\text{Sw}\left(x,y,a_3,z\right)$ beam can then be described in terms of again an optical swallowtail catastrophe $\text{Sw}\left(A_1,A_2,A_3\right)$, whose control parameters are functions of $x,y,a_3,z$ in the form of:
\small
\begin{eqnarray}
\label{eq:SwA3Propagation}
\text{Sw}(x,y,a_3,z) &=& \exp\left[\text{i}\psi\right] \cdot \text{Sw}\left(A_1, A_2, A_3\right), \text{where} \\
\nonumber
A_1 &=& A_1(x,y,a_3,z),\\
\nonumber
A_2 &=& A_2(y,a_3,z),\\
\nonumber
A_3 &=& A_3(a_3,z),\\
\nonumber
\psi &=& \psi(x,y,a_3,z).
\end{eqnarray}
\normalsize
The complete analytical expression for the dynamics are given in the SM.

\begin{figure}[]
\centering
\includegraphics[width=\columnwidth]{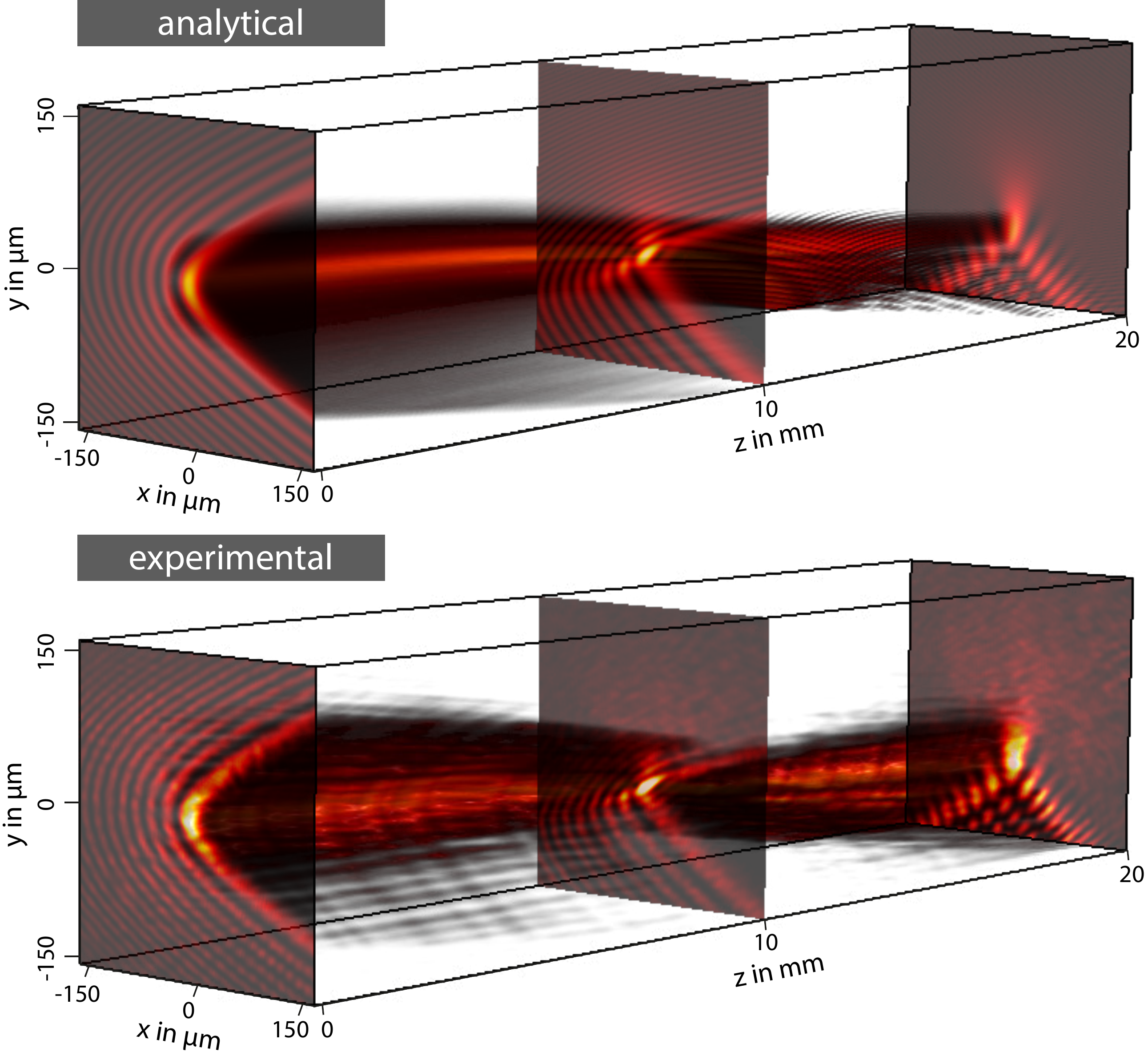}%
\caption{Propagation of the $\text{Sw}\left(x,y,0\right)$ beam. Top: Intensity volume evaluated according to \eqref{eq:SwA3Propagation}. Bottom: Experimentally obtained intensity volume with same parameters.}%
\label{fig:SwA3Prop}%
\end{figure}

Plotting \eqref{eq:SwA3Propagation} for different $z$-positions allows visualizing the dynamics of a $\text{Sw}\left(x,y,a_3\right)$ beam. We set $a_3 = 0$. The analytically calculated propagation is depicted in Fig.~\ref{fig:SwA3Prop}, top. Furthermore, we obtain the volume intensity distribution of the beam experimentally for different $z$-positions and show them in Fig.~\ref{fig:SwA3Prop}, bottom. We chose transverse feature sizes of $x_0 = y_0 = \unit[8]{\upmu m}$ and a propagation distance of $\unit[20]{mm}$, which covers the most interesting effects in the sketched intensity volume.

Analysis of \eqref{eq:SwA3Propagation} reveals that at $z = \unit[0]{mm}$ the $\text{Sw}\left(x,y,a_3\right)$ beam shows mirror symmetry with respect to $y$:
\begin{equation}
\text{Sw}\left(x, y, a_3, z\right) = \text{Sw}^*\left(x, -y, a_3, -z\right),
\label{eq:SwA3Symmetry}
\end{equation}
where the asterisk denotes the complex conjugation. Due to the symmetry, the beam's propagation for negative $z$-values is not shown, but has also been observed numerically and experimentally.

A unique feature of the $\text{Sw}\left(x,y,a_3\right)$ beam becomes apparent: During propagation, the transverse field at the origin ($z = \unit[0]{mm}$) turns out to consist of a \textit{fast diffracting contribution} that quickly vanishes due to its transverse momentum (e.g.~visible at $z \approx \unit[10]{mm}$), subsequently revealing a field 	contribution (remaining field at $z \approx \unit[20]{mm}$) that resembles a Pearcey beam in the field distribution as well as in the propagation. 

This similarity is additionally manifested in the distribution of Fourier components, since they are located on a parabola for both, the Pearcey and $\text{Sw}\left(x,y,a_3\right)$ beam~\cite{Ring2012, Zannotti2017}: 
\begin{equation}
\label{eq:SwallowtailA3Pearcey}
\tilde{\text{Sw}}\left(k_x,k_y,0\right) = \tilde{\text{Pe}}\left(k_x,k_y\right) \cdot e^{\text{i}\left(x_{0}k_x\right)^5} e^{-\text{i}\left(x_{0}k_x\right)^4}.
\end{equation}
Due to this spectral similarity, we express the $\text{Sw}\left(x,y,a_3 = 0\right)$ beam as a convolution of a Pearcey beam with a dimensionally reduced swallowtail beam:
\begin{equation}
\begin{split}
\label{eq:SwallowtailA3Convolution}
\text{Sw}\left(x,y,0\right) &= \text{Pe}\left(x,y\right) \ast \\
&\frac{\delta(y)}{x_{0}} e^{\text{i}\left(\frac{x}{5x_{0}} - \frac{4}{3125}\right)} \text{Sw}\left(\frac{x}{x_{0}} - \frac{3}{125}, - \frac{4}{25}, -\frac{2}{5}\right).
\end{split}
\end{equation}
Here, $\delta(...)$ is the delta-function. Note that the initial beam profile of the artificially designed $\text{Sw}\left(x,y,a_3\right)$ beam shows a swallowtail caustic as the mapping of a cross-section through the higher-dimensional 3D control parameter space to the lower-dimensional 2D transverse plane. Due to this, the swallowtail structure in Fig.~\ref{fig:SwA3Prop} becomes unstable during propagation with respect to the otherwise geometrically stable swallowtail catastrophe in nonlinear systems, and decays to a lower-dimensional optical cusp catastrophe. We will discuss the dynamics of the caustic in more detail in section \ref{sc:SwCaustic}.

\section{Dynamic transformation of swallowtail to butterfly caustics}

The dynamics of the diverse optical swallowtail beams and their caustics are fundamentally different if compared with each other. In order to demonstrate that caustic beams of a certain oder $n$ may show structural similarities with higher-order caustics during propagation, we exemplary investigate a $\text{Sw}\left(x,a_2,y\right)$ beam, thus set $a_2 = $ const.~and chose $\alpha = 1$, $\beta = 3$. Applying \eqref{eq:CausticPropagation} and subsequently performing a Tschirnhaus transform with appropriate substitutions, we express the propagation of the $\text{Sw}\left(x,a_2,y,z\right)$ beam in terms of the higher-order static butterfly beam $\text{Bu}(B_1,B_2,B_3,B_4)$ with an overall phase factor $\psi$ in dependence of new control parameters $\mathbf{B} = \{B_1,B_2,B_3,B_4\}$ as functions of $x,y,z,a_2$:
\small
\begin{eqnarray}
\label{eq:SwA2Propagation}
\text{Sw}(x,a_2,y,z) &=& \gamma^{-1}\exp\left[\text{i}\psi\right] \cdot \text{Bu}\left(B_1, B_2, B_3, B_4\right), \qquad\\
\nonumber 
&&\text{where}\\
\nonumber
B_1 &=& B_1(x,a_2,y,z);\\
\nonumber
B_2 &=& B_2(a_2,y,z);\\
\nonumber
B_3 &=& B_3(y,z);\\
\nonumber
B_4 &=& B_4(z);\\
\nonumber
\psi &=& \psi(x,a_2,y,z);\\
\nonumber
\end{eqnarray}
\normalsize
The explicit functions for the new control parameters are given in the SM. The butterfly beam is defined as $\text{Bu}\left(a_1, a_2, a_3, a_4\right) = C_{6}\left(a_1, a_2, a_3, a_4\right)$. We have condensed $\gamma = \left(-z/z_{ey}\right)^{1/6}$, and used characteristic lengths $\mathbf{z_e}$ as defined previously. We note that the propagation of a $\text{Sw}\left(a_1,x,y\right)$ beam can be expressed in a similar way in terms of static butterfly beams by following \eqref{eq:CausticPropagation}.

\begin{figure}[]
\centering
\includegraphics[width=\columnwidth]{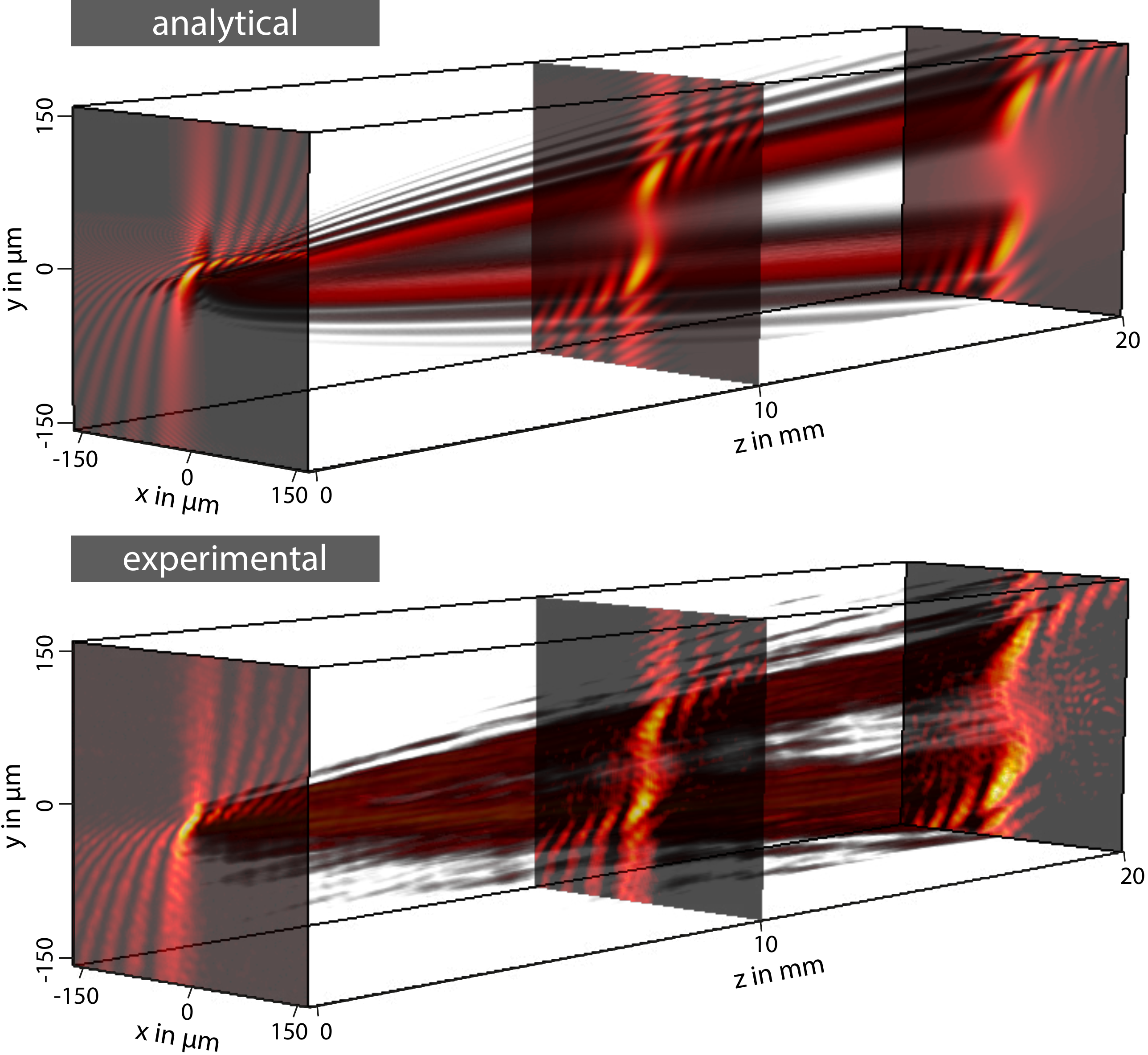}%
\caption{Propagation of the $\text{Sw}\left(x,0,y\right)$ beam. Top: Intensity volume evaluated according to \eqref{eq:SwA2Propagation}. Bottom: Experimentally obtained intensity volume with same parameters.}%
\label{fig:SwA2Prop}%
\end{figure}
\eqref{eq:SwA2Propagation} is plotted for different $z$-values and illustrated in Fig.~\ref{fig:SwA2Prop}, top. We obtained the dynamics as well in the experiment, which is shown below. The transverse dimensions are $x_0 = y_0 = \unit[8]{\upmu m}$ and the propagation distance is $\unit[20]{mm}$. Due to the point-symmetry of the $\text{Sw}\left(x,a_2,y\right)$ beam (cf. \eqref{eq:SwA2Propagation})
\begin{equation}
\text{Sw}\left(x, a_2, y, z\right) = -\text{Sw}^*\left(x, -a_2, y, -z\right),
\label{eq:SwA2Symmetry}
\end{equation}
and for illustrative reasons, we restrict the evaluation to $z \leq 0$. The asterisk denotes the complex conjugation.

The plotted analytical description of the propagation of a $\text{Sw}\left(x,a_2,y\right)$ beam is in high agreement with the corresponding experimental realization. The transverse plane at $z = \unit[0]{mm}$ shows no mirror nor point symmetry, which can be seen in Fig.~\ref{fig:SwA2Prop} and be proved by considering \eqref{eq:CatastropheIntegral}. However, during propagation the structure separates into two individual structures, showing point symmetry. This suggests that we can assume the $\text{Sw}\left(x,a_2,y\right)$ beam at the origin to consist of two major contributions: One fast diffracting part that breaks the symmetry in the initial plane, and a second one that remains during propagation and exhibits point symmetry. After a certain propagation distance ($z \approx \unit[10]{mm}$), the emerging point symmetric structure can easily be considered as a doubled Pearcey-beam-like structure. Nevertheless, the contribution rather occurs due to the similarity of the $\text{Sw}\left(x,a_2,y\right)$ beam with a $\text{Bu}\left(x,a_2,y,a_4\right)$ beam, whose $z = \unit[0]{mm}$ real space appearance~\cite{Zannotti2017} resembles the field distribution of the $\text{Sw}\left(x,a_2,y\right)$ beam after a certain propagation distance. Their spectra both correspond to cubic functions, thus 
\begin{equation}
\tilde{\text{Sw}}\left(x,a_2,y\right) = \tilde{\text{Bu}}\left(x,a_2,y,a_4\right) \cdot e^{-\text{i}\left[(x_{0}k_x)^6 - (x_{0}k_x)^5 + (x_{0}k_x)^4\right]}.
\label{eq:ButterflySwallowtail}
\end{equation}
This leads to the real space expression
\begin{equation}
\begin{split}
\label{eq:SwallowtailA2Convolution}
\text{Bu}\left(x,0,y,0\right) &= \text{Sw}\left(x,0,y\right) \ast \\
&\frac{\delta(y)}{x_{0}} e^{\text{i}\left(\frac{x}{6x_{0}} - \frac{31}{46656}\right)} \text{Bu}\left(\frac{x}{x_{0}} + \frac{5}{324}, \frac{19}{144}, \frac{13}{27}, \frac{7}{12}\right),
\end{split}
\end{equation}
which indicates an inverse convolution. Properties of the static $\text{Sw}\left(x,a_2,y\right)$ beam at the origin are mainly determined by the characteristics of the $\text{Bu}\left(x,a_2,y,a_4\right)$ beam, therefore revealing similar field distributions during propagation as the $\text{Bu}\left(x,a_2,y,a_4\right)$ beam (cf.~\cite{Zannotti2017}).

\section{Dynamic decay of the swallowtail caustic to a lower-order cusp}
\label{sc:SwCaustic}

\begin{figure}[!b]
\centering
\includegraphics[width=1\columnwidth]{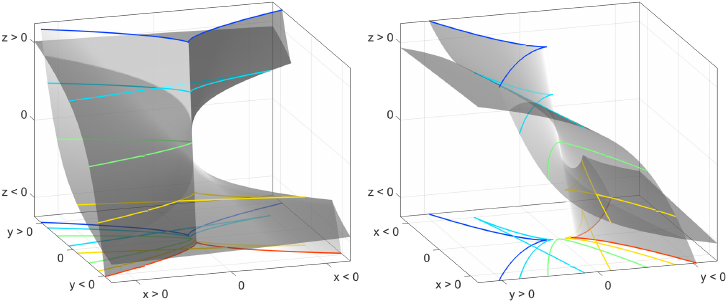}%
\caption{Dynamic of the swallowtail caustic. Two different perspectives to the 3D caustic surface during propagation. The initial swallowtail caustic $(z=0)$ decays to cusps $(z \neq 0)$. Representative $z$ positions are highlighted with different colours and mapped as contours to the $x$-$y$-plane.}
\label{fig:PropagationOfSwCaustics}%
\end{figure} 	

The propagation of the $\text{Sw}\left(x,y,a_3\right)$ beam shown in Fig.~\ref{fig:SwA3Prop} gives rise to further study the dynamics of the corresponding caustic. By parametrizing \eqref{eq:SwA3Propagation} and analyzing the structure of the caustics according to \eqref{eq:DegeneratedCriticalPoints}, we find the $z$-dependent dynamics of the caustic in the transverse plane. 

Fig.~\ref{fig:PropagationOfSwCaustics} illustrates the parametrized surface of the caustic in real space $(x,y,z)$. The plane at $z = 0$, where the green line highlights the form of the caustic, equals the front plane at Fig.~\ref{fig:SwA3Prop}.

The surface appears to consist of two intertwined slices, merging point symmetrically at the origin. Each of them has a cuspoid form that changes with the $z$-position, becomes flattened and is strongly bent. This can be recognized as the \textit{fast diffracting contribution} described above. The remaining part is the cusp of the respective other slice. According to \eqref{eq:SwA3Symmetry}, we find that the previously discussed properties of the beam resemble the dynamics of the caustic. Since the lower-dimensional initial light field contains the fingerprint of the optical swallowtail caustic due to the mapping of a cross-section from the higher-dimensional control parameter space, the expected decay of the swallowtail catastrophe to a lower-order cusp is apparent. Though the structural stability of the swallowtail catastrophe with a 3D control parameter space is lost in 2D, the demonstrated swallowtail light fields show caustic structures that exhibit novel and unique propagation properties like e.g.~high-intensities on curved paths.

\section{Conclusion}

Perturbing the external control parameters of a nonlinear system described by a potential function leads to so-called catastrophes at locations where bifurcations suddenly shift~\cite{Thom1975, Arnold2003, Zeeman1979}. These singularities of the gradient map of potential functions manifest as caustics in light, and where extensively studied as natural phenomena in the late seventies and eighties~\cite{Berry1976, Berry1979, Berry1980, Berry1975, Kravtsov1983}. Mapping catastrophes to light fields via the paraxial catastrophe integral of \eqref{eq:CatastropheIntegral} is a well-known approach. However, it took until 2007 that spatial light modulators were used to create Airy beams~\cite{Siviloglou2007a}. This allows for the first time the controlled mapping of caustics to paraxial light, and starts the renaissance of designing artificial caustic beams. 

In our contribution, we considered the complete class of caustic beams depending on a single state parameter $s$ to map higher-order catastrophes to the lower-dimensional initial transverse plane of paraxial beams and derived analytically a general equation for their Fourier spectra (SM) and propagation. Our approach connects two of the control parameters $\mathbf{a}$ with the transverse spatial coordinates $(x,y)^T$, thus the caustic beams are realized as cross-sections through the higher-dimensional control parameter space. We showed that, depending on the control parameters identified with the spatial coordinates, the propagation of caustic beams of order $n$ can be calculated in terms of higher-order static caustic beams of order $2(n-2)$.

We demonstrated this by analytically calculating and experimentally obtaining the dynamics of the $\text{Sw}\left(x,y,a_3\right)$ and $\text{Sw}\left(x,a_2,y\right)$ beams. We proved that the evolution of the latter beam is linked to a static butterfly beam. Furthermore, we analyzed the dynamics of the swallowtail caustic and showed how its surface evolves in the 3D real space. We thereby demonstrated that the swallowtail catastrophe at an initial plane continuously decays to a lower-order cusp catastrophe during propagation. 

The demonstrated optical catastrophes are highly attractive for microscopy and super-resolution applications. The propagation of their high-intensity rims near the caustics capable to form tailored structures are unique for each order of catastrophe and parameter set and pave the way towards advanced micro-machining on tailored curves~\cite{Mathis2012} and the realization of waveguides with a rich diversity of light guiding paths~\cite{Zannotti2017a}. 


\newpage
\phantom{.}
\newpage

\setcounter{subsection}{0}
\setcounter{section}{0}
\setcounter{equation}{0}
\setcounter{page}{1}

\onecolumn

\begin{center}
\huge{Dynamics of the optical swallowtail catastrophe \Large Supplementary material\normalsize}
\end{center}

\section{Fourier transform of caustic beams}

We perform a 2D Fourier transform with respect to two control parameters $a_\alpha,~a_\beta$. The $n-2$ dimensional control parameter space $\mathbf{a}$ is spanned by the coordinates $a_j$, where $j = 1,...,n-2$. We introduce $k_\alpha, k_\beta$ as Fourier frequencies of $a_\alpha, a_\beta$, where $\alpha, \beta \in \left[1,...,n-2\right]$. With $\mathbf{a}'$ we denote all remaining control parameters $\mathbf{a}$ except $a_\alpha,~a_\beta$. We neglected $2\pi$ scaling factors for reasons of clarity.
\begin{eqnarray}
\label{eq:CausticFourier1}
\begin{split}
&\tilde{C}_n\left(k_\alpha,k_\beta,\mathbf{a}'\right) = \int\limits_{\mathds{R}^2} C_n(\mathbf{a}) e^{-\text{i}k_\alpha a_\alpha} e^{-\text{i}k_\beta a_\beta} \mathrm{d}a_\alpha \mathrm{d}a_\beta\\
&= \delta_{k_\alpha,r^+} \cdot \delta\left(\left|a_{0\alpha} k_\alpha\right|^\frac{\beta}{\alpha} - a_{0\beta}k_\beta\right) \cdot \frac{a_{0\alpha}a_{0\beta}}{\left|\alpha\left(a_{0\alpha}k_\alpha\right)^{\frac{\alpha-1}{\alpha}}\right|} \\
&\cdot \exp \left[ \text{i} \left( \left|a_{0\alpha} k_\alpha\right|^\frac{n}{\alpha} + \sum\limits_{\substack{j = 1 \\ j \neq \alpha, \beta}}^{n-2}\frac{a_j}{a_{0j}}\left|a_{0\alpha} k_\alpha\right|^{\frac{j}{\alpha}} \right) \right] \\
&+ \left(\delta_{\frac{\alpha}{2},h} \cdot \delta_{k_\alpha,r^+} + \delta_{\frac{\alpha + 1}{2},h} \cdot \delta_{-k_\alpha,r^+} \right)\\
&\cdot \delta\left(\left(- \left|a_{0\alpha} k_\alpha\right|^\frac{1}{\alpha}\right)^\beta - a_{0\beta}k_\beta\right) \cdot \frac{a_{0\alpha}a_{0\beta}}{\left|\alpha\left(a_{0\alpha}k_\alpha\right)^{\frac{\alpha-1}{\alpha}}\right|} \\
\label{eq:CausticFourier2}
&\cdot \exp \left[ \text{i} \left( \left(-\left|a_{0\alpha} k_\alpha\right|^\frac{1}{\alpha}\right)^n + \sum\limits_{\substack{j = 1 \\ j \neq \alpha, \beta}}^{n-2} \frac{a_j}{a_{0j}} \left(- \left|a_{0\alpha} k_\alpha\right|^{\frac{1}{\alpha}}\right)^j \right) \right]
\end{split}
\end{eqnarray}
\normalsize
Thereby, $h \in \mathds{Z}$ and $r^+ \in \mathds{R}_0^+$. The Kronecker delta $\delta_{i,j}$ equals $1$ if $i = j$ and equals $0$ for $i \neq j$. The choice whether $\alpha$ is even or odd is crucial. Therefore, the term with $\delta_{\frac{\alpha}{2},h}$ contributes only if $\alpha$ is even, while the term that depends on $\delta_{\frac{\alpha + 1}{2},h}$ contributes only if $\alpha$ is odd. Additionally, $\delta_{k_\alpha,r^+}$ only allows positive $k_\alpha$ to contribute in corresponding terms, whereas $\delta_{-k_\alpha,r^+}$ selects only negative $k_\alpha$. 

Thus, the Kronecker deltas and delta functions in \eqref{eq:CausticFourier2} determine which quadrants of the Fourier space are occupied by Fourier components and which distribution they obey, may it be polynomial or an expression with rational exponent.

\section{Analytical expressions for the dynamics of the optical swallowtail catastrophes}

With Eq.~(3) in the main document, it is a straight forward task to calculate analytically the propagation of any order of optical catastrophe that was mapped to the 2D initial transverse field via Eq.~(1). Since we investigate here the swallowtail catastrophe, $n = 5$, with co-dimension $n-2 = 3$. For all three cases, after applying Eq.~(3), a Tschirnhaus transform is necessary yielding the canonical expression of the potential function $P_n$, which in turn allows identifying the potential with a generic catastrophe. The dynamics of the three swallowtail beams are given in the following.

\subsection{Propagation of the $\text{Sw}\left(x,y,a_3\right)$ beam}

Here, $\alpha = 1$ and $\beta = 2$, or vice versa, since $a_3 = \text{const.}$.
\begin{eqnarray}
\label{eq:SwA3Propagation}
\text{Sw}(x,y,a_3,z) &=& \exp\left[\text{i}\psi\right] \cdot \text{Sw}\left(A_1, A_2, A_3\right), \quad \text{where} \\
\nonumber
A_1 &=& \frac{x}{x_{0}} -\frac{3}{125}\left(\frac{z}{z_{ey}}\right)^4 - \frac{2}{5}\left(\frac{z^2}{z_{ex}z_{ey}}\right) \\
\nonumber
&&+ \frac{3}{25} \frac{a_3}{a_{03}}\left(\frac{z}{z_{ey}}\right)^2 + \frac{2}{5}\frac{y}{y_{0}} \left(\frac{z}{z_{ey}}\right),\\
\nonumber
A_2 &=& \frac{y}{y_{0}} - \frac{4}{25}\left(\frac{z}{z_{ey}}\right)^3 +\frac{3}{5}\frac{a_3}{a_{03}}\left(\frac{z}{z_{ey}}\right) - \left(\frac{z}{z_{ex}}\right),\\
\nonumber
A_3 &=& \frac{a_3}{a_{03}} - \frac{2}{5}\left(\frac{z}{z_{ey}}\right)^2, \qquad \text{and} \\
\nonumber
\psi &=& -\frac{4}{3125}\left(\frac{z}{z_{ey}}\right)^5 - \frac{1}{25}\left(\frac{z^3}{z_{ex}z_{ey}^2}\right) \\
\nonumber
&&+ \frac{1}{125}\frac{a_3}{a_{03}}\left(\frac{z}{z_{ey}}\right)^3 +\frac{1}{25}\frac{y}{y_{0}}\left(\frac{z}{z_{ey}}\right)^2 \\
\nonumber
&&+ \frac{1}{5}\frac{x}{x_{0}}\left(\frac{z}{z_{ey}}\right).
\end{eqnarray}
\normalsize

\subsection{Propagation of the $\text{Sw}\left(x,a_2,y\right)$ beam}

Here, $\alpha = 1$ and $\beta = 3$, or vice versa, since $a_2 = \text{const.}$.

\begin{eqnarray}
\label{eq:SwA2Propagation}
\text{Sw}(x,a_2,y,z) &=& \frac{1}{\gamma}\exp\left[\text{i}\psi\right] \cdot \text{Bu}\left(B_1, B_2, B_3, B_4\right), \quad \text{where}\\
\nonumber
B_1 &=& \frac{1}{\gamma}\left( \frac{x}{x_{0}} + 2\left(\frac{z}{z_{ex}} - \frac{a_2}{a_{02}}\right)h + 3\frac{y}{y_{0}} h^2 + 4 h^4  \right),\\
\nonumber
B_2 &=& \frac{1}{\gamma^2}\left( \frac{a_2}{a_{02}} -\frac{z}{z_{ex}} + 3\frac{y}{y_{0}} h + \frac{15}{2} h^3 \right),\\
\nonumber
B_3 &=& \frac{1}{\gamma^3}\left(\frac{y}{y_{0}} + \frac{20}{3} h^2\right),\\
\nonumber
B_4 &=& -\frac{1}{\gamma^4}\frac{5}{2}h, \qquad \text{and} \\
\nonumber
\psi &=& -\frac{x}{x_{0}} h +  \left(\frac{z}{z_{ex}} - \frac{a_2}{a_{02}}\right)h^2 + \frac{y}{y_{0}} h^3 + 30 h^5,
\nonumber
\end{eqnarray}
\normalsize

with $\gamma = \left(-\frac{z}{z_{ey}}\right)^{\frac{1}{6}}$, and $h = \frac{1}{6\gamma^6}$.

\subsection{Propagation of the $\text{Sw}\left(a_1,x,y\right)$ beam}

In the main part of this work, we demonstrated the propagation of the $\text{Sw}\left(x,y,a_3\right)$ and $\text{Sw}\left(x,a_2,y\right)$ beam, with particular emphasis on the fact that the $\text{Sw}\left(x,y,a_3,z\right)$ beam can still be described in terms of a swallowtail beam, whereas for other choices of $\alpha$ and $\beta$, as it is the case for the $\text{Sw}\left(x,a_2,y,z\right)$ and $\text{Sw}\left(a_1,x,y,z\right)$ beams, their propagation is described in terms of higher-order butterfly beams.

Here we add the remaining analytical equation that describes the propagation of the $\text{Sw}\left(a_1,x,y\right)$ beam. Thus, we chose $\alpha = 2$ and $\beta = 3$, or vice versa, since $a_1 = \text{const.}$.

\begin{eqnarray}
\label{eq:SwA1Propagation}
&&\text{Sw}(a_1,x,y,z)  = \frac{1}{\gamma}\exp\left[\text{i}\psi\right] \times \text{Bu}\left(B_1, B_2, B_3, B_4\right), \quad \text{where} \qquad\\
&&B_1 = \frac{1}{\gamma}\left( \frac{a_1}{a_{01}} -2 \frac{x}{x_0}h - 3\frac{y}{y_0}h^2 - \frac{2}{3}\frac{z_{ey}}{z_{ex}}h^2 + 4h^4  \right),\nonumber\\
&&B_2 = \frac{1}{\gamma^2}\left( \frac{x}{x_0} - 3\frac{y}{y_0}h + \frac{z_{ey}}{z_{ex}}h -\frac{15}{2}h^3 \right),\nonumber\\
&&B_3 = \frac{1}{\gamma^3}\left( \frac{y}{y_0} - \frac{2}{3}\frac{z_{ey}}{z_{ex}} + \frac{20}{3}h^2 \right),\nonumber\\
&&B_4 = -\frac{1}{\gamma^4}\left( \frac{z_{ey}}{z_{ex}}\gamma^6 - \frac{15}{6}h \right), \qquad \text{and} \nonumber\\
&&\psi = - \frac{a_1}{a{_01}}h + \frac{x}{x_0}h^2 -\frac{y}{y_0}h^3 - 6\frac{z_{ey}}{z_{ex}}h^3 - 30h^5, \nonumber
\end{eqnarray}
\normalsize
with $\gamma = \left(-\frac{z}{z_{ey}}\right)^{\frac{1}{6}}$, and $h = \frac{1}{6\gamma^6}$.

\end{document}